\documentclass[prl,twocolumn,nofootinbib,floatfix]{revtex4} 
\usepackage{graphicx}
\ifnum\lefthyphenmin<2\lefthyphenmin=2\fi
\ifnum\righthyphenmin<2\righthyphenmin=2\fi

\newcommand{\cF}{{\cal F}}
\newcommand{\cS}{{\cal S}}
\newcommand{\verylongarrow}{\setlength{\unitlength}{1cm} 
  \,\begin{picture}(2,0.1)\vector(1,0){2}\end{picture}\,}

\begin{document}
\title{Translocation of structured polynucleotides through nanopores}
\author{Ulrich Gerland$^*$\footnote[2]{Present address: 
Physics Department, Ludwig-Maximilians-Universit\"at, Theresienstr. 37, 
80333 M\"unchen, Germany. Email: Ulrich.Gerland@physik.uni-muenchen.de}}
\author{Ralf Bundschuh$^\ddagger$}
\author{Terence Hwa$^*$}
\affiliation{$^*$Department of Physics and Center for Theoretical Biological 
Physics, University of California at San Diego, La Jolla, 
California 92093-0319}
\affiliation{$^\ddagger$Department of Physics, The Ohio State University,
174 W 18th Av., Columbus, Ohio 43210-1106}
\date{\today}
%
%
\begin{abstract}
We investigate theoretically the translocation of structured RNA/DNA 
molecules through narrow pores which allow single but not double strands 
to pass. 
The unzipping of basepaired regions within the molecules presents 
significant kinetic barriers for the translocation process. 
We show that this circumstance may be exploited to determine 
the full basepairing pattern of polynucleotides, including RNA pseudoknots. 
The crucial requirement is that the translocation dynamics (i.e., the
length of the translocated molecular segment) needs to be recorded as a 
function of time with a spatial resolution of a few nucleotides. 
This could be achieved, for instance, by applying a mechanical driving force 
for translocation and recording force-extension curves (FEC's) with a device 
such as an atomic force microscope or optical tweezers. 
Our analysis suggests that with this added spatial resolution, nanopores
could be transformed into a powerful experimental tool to study the 
folding of nucleic acids. 
\end{abstract}
\maketitle
A series of recent experiments studied the translocation of DNA and 
RNA molecules through narrow pores, which allow single but not double 
strands to pass \cite{Kasianowicz96,Akeson99,Meller00,Henrickson00,
Vercoutere01,Meller01,Bates03,Sauer-Budge03}, see Ref.~\cite{Meller03} 
for a review. 
These investigations pursued two main goals: 
(i) to probe in a well-defined model system the physics of biopolymer 
translocation across membranes, a process which is ubiquitous in cell 
biology, and (ii) to explore the potential of nanopores as a 
single-molecule tool. 
In the experiments so far, a membrane protein, $\alpha$-hemolysin, was used 
as the pore. 
An electric field acting on the negatively charged DNA/RNA backbone 
drives the molecules through the pore, and 
translocation is monitored by measuring the induced ionic current, 
which is strongly reduced while a DNA/RNA chain blocks the pore.  
Until very recently \cite{Vercoutere01,Sauer-Budge03}, the experiments 
have focused on the translocation of unstructured, mostly homopolymeric 
molecules, a problem which has also received considerable theoretical 
interest \cite{Sung96,DiMarzio97,Muthukumar99,Lubensky99,Muthukumar01,
Chuang01,Ambjornsson02,Flomenbom03,Metzler03}. 
For such unstructured molecules, the main results regarding the 
above goals were that 
(i) the basic physics of translocation is adequately described by a 
drift-diffusion process, in which monomers hop randomly in and out of the 
pore with a directional bias due to the applied voltage \cite{Lubensky99}, 
and (ii) nanopores could possibly be developed into rapid sequencing 
devices, since the ionic current during blockage displays a weak 
sequence-dependence \cite{Akeson99,Meller00}. 

In contrast, for structured polynucleotides, both the basic physics and 
the potential applications of translocation still remain largely unexplored. 
Experimentally, important first steps have been taken by studying the 
translocation of simple hairpin (i.e., stem-loop) structures 
\cite{Vercoutere01} and the unzipping of double-stranded DNA through a 
nanopore \cite{Sauer-Budge03}. 
However, a general theoretical framework to describe translocation of 
these as well as more complex RNA/DNA structures is currently lacking. 
Here, we first construct such a framework and then use it to investigate 
the potential of nanopores as a single-molecule tool for the study of 
biopolymer folding. 

\begin{figure}[b]
  \includegraphics[width=7.5cm]{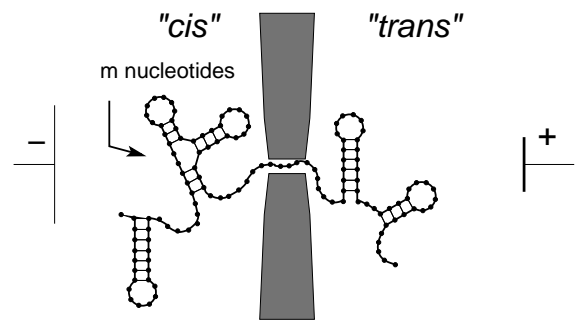}
  \caption{Sketch of a structured polynucleotide that is driven across a 
  nanopore which allows single but not double strands to pass. Here, the 
  driving force causing translocation from the {\it cis} to the {\it trans} 
  side is exerted by an electric field that acts on the negatively charged 
  backbone of the molecule.}
  \label{sketch_pore} 
\end{figure} 

In this article, we are interested in the generic physical aspects of the 
translocation process that neither depend on the specific properties of 
a particular protein pore, nor on the detailed way in which the driving 
force for translocation is applied. 
As in previous theoretical studies \cite{Sung96,DiMarzio97,Muthukumar99,
Lubensky99,Muthukumar01,Chuang01,Ambjornsson02}, we use a 
coarse-grained model which treats the pore basically as a separator between 
a {\it cis} and a {\it trans} part of the molecule with a characteristic 
friction coefficient, see the sketch in Fig.~\ref{sketch_pore}. 
Presumably this description will apply directly to solid-state nanopores 
\cite{Li01,Storm03}, which can now be fabricated with sizes down to 
$\sim2$~nm, not much larger than the $\sim1.5$~nm aperture of the 
$\alpha$-hemolysin pore and slightly smaller than the $\sim2.2$~nm 
diameter of double-stranded DNA or stems in RNA.  
Also, we do not consider the full three-dimensional (tertiary) structure 
of the molecules, but focus on the basepairing pattern, i.e. the 
secondary structure including possible pseudoknots, which are the only 
structural features present when there are no divalent metal ions in 
the solution. 
Unless stated otherwise, the term `structure' refers here to this 
basepairing pattern. 
While both our theoretical framework and our conclusions apply equally 
to RNA and single-stranded DNA, the RNA case is particularly interesting, 
since structured RNA's have a multitude of functions in molecular biology 
and RNA folding is an active field of research 
\cite{Cech93,Brion97,Tinoco99,Thirumalai01}.

\section*{General theoretical framework}
Fig.~\ref{sketch_pore} depicts schematically the driven translocation of a 
structured polynucleotide from the {\it cis} to the {\it trans} side of 
the pore. 
We seek here a convenient reduced description of this translocation process, 
rather than modeling the full three-dimensional polymer dynamics explicitly. 
Our approach is similar in spirit to the existing models for the case of 
unstructured polymers \cite{Sung96,Muthukumar99,Lubensky99}, where the 
translocation dynamics is formulated in terms of a single variable, e.g. 
the number of nucleotides, $m$, on the {\it cis} side, see 
Fig.~\ref{sketch_pore}. 
The dynamics, $m(t)$, is stochastic and can be described by `hopping rates', 
$k_-(m)$ and $k_+(m)$, for forward and backward motion of the nucleotide 
chain through the pore with a stepsize of one monomer. 
The external force on the molecule leads to an imbalance in the hopping 
rates, $k_-(m)>k_+(m)$, and hence a mean drift towards the {\it trans} side. 
For unstructured molecules the one-dimensional description is 
permissible, if the relaxation of the polymer degrees of freedom on both 
sides of the pore is faster than the hopping process. 
This assumption does not hold for arbitrarily long polymers, since the 
relaxation time increases with the polymer length \cite{Lubensky99,Chuang01}, 
however for lengths on the order of a thousand bases, the 
one-dimensional description is adequate under typical experimental 
conditions \cite{Lubensky99}. 
The residual effect of the polymer ends is then only to introduce an 
entropic barrier for translocation, which leads to a weak $m$-dependence 
of the hopping rates. 

For structured molecules, the translocation dynamics is considerably more 
complicated, since the dynamics of the `reaction coordinate', $m(t)$, is 
then coupled to the dynamics of the basepairing patterns on both sides: 
the structure on the {\it cis} side, $\cS_{cis}(t)$, affects the forward 
rate, while the structure on the {\it trans} side, $\cS_{trans}(t)$, affects 
the backward rate, 
\begin{eqnarray}
  \label{hopping}
  m &\stackrel{k_-(m,\,\cS_{cis}(t))}{\verylongarrow}& m-1 
  \nonumber \\[0.2cm]
  m &\stackrel{k_+(m,\,\cS_{trans}(t))}{\verylongarrow}& m+1 \;.
\end{eqnarray}
In two limiting cases however, the process can be modeled by 
a one-dimensional Brownian walk as for unstructured molecules, but 
with a complex sequence/structure-dependent free energy landscape 
$\cF(m)$ along the coordinate $m$: 
(A) If the dynamics of the basepairing patterns $\cS_{cis}(t)$ and 
$\cS_{trans}(t)$ is much faster than the hopping process, the landscape 
is determined by the ensemble free energy of all basepairing patterns on 
the {\it cis} and {\it trans} side. 
(B) In the opposite limit, the basepairing pattern on the {\it cis} side 
is essentially frozen and is unzipped basepair by basepair as it is driven 
through the pore. The landscape is then determined by the basepairing 
energetics of the particular molecular structure prior to translocation, 
see below. 
In both cases, the free energy naturally decomposes into three parts, 
\begin{equation}
  \label{general_FE}
  \cF(m) = \cF_{cis}(m) + \cF_{trans}(m) + \cF_{\rm ext}(m) \;,
\end{equation}
where $\cF_{cis}(m)$ and $\cF_{trans}(m)$ denote the intrinsic binding free 
energies of the {\it cis} and {\it trans} parts of the molecule, while 
$\cF_{\rm ext}(m)$ describes the effect of the external force. 
Given $\cF(m)$, the simplest form for the hopping rates $k_{\pm}(m)$ 
which satisfies the detailed balance condition 
$k_+(m)/k_-(m+1)=e^{-\beta[\cF(m+1)-\cF(m)]}$ 
(with $\beta=1/k_BT$) is 
\begin{equation}
  \label{hopping_rate}
  k_{\pm}(m)=k_0\,e^{-\beta\cdot{\rm max}\{\cF(m\pm 1)-\cF(m)\,,
  \,0\}}\;. 
\end{equation}
Here, $k_0$ denotes a microscopic rate constant, which can in principle 
be tuned by adjusting the properties of the pore. It can be interpreted as 
a friction coefficient and corresponds approximately to the bare hopping 
rate for unstructured molecules at zero external force (typical experimental 
estimates for $k_0$ are on the order of $10^5\,{\rm s}^{-1}$ \cite{Meller01}). 
The dynamics of the translocation process, as described by 
Eqs.~(\ref{general_FE}) and (\ref{hopping_rate}) is dominated by 
energetic barriers due to basepairing, whereas the above-mentioned entropic 
barrier is completely negligible for structured molecules. 
These energetic barriers lead to arrests during translocation, as clearly 
observed already in the experiments with simple hairpins \cite{Vercoutere01} 
and double-stranded DNA \cite{Sauer-Budge03}.

\begin{figure}[t]
\includegraphics[width=8.5cm]{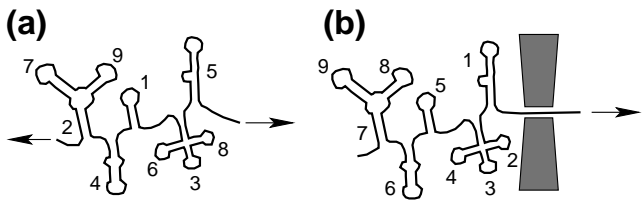}
\caption{Unzipping a structured molecule by pulling on its ends is 
fundamentally different from unzipping by driven translocation through a 
narrow pore. (a) For pulling on the ends, the stems (i.e., contiguously 
basepaired segments) in the molecule unfold in an order determined by 
their relative stability and the topology of the structure (a possible 
order 1--9 is indicated). (b) In contrast, the pore forces the stems to 
unfold in a linear order along the sequence, as again indicated by the 
numbering 1--9. 
} 
\label{comparison} 
\end{figure} 

\section*{Pulling through a pore}
\noindent {\bf Qualitative aspects.} 
We now make use of the theoretical framework constructed above to investigate 
which information on structured molecules could be derived from pore 
translocation experiments. To this end, 
it is useful to compare unzipping by driven translocation through a 
nanopore with the more conventional way of unzipping by applying a force 
on the ends of a biopolymer, see e.g. \cite{Rief97,Liphardt01,Onoa03}. 
As illustrated in Fig.~\ref{comparison}, the two approaches differ 
fundamentally: Pulling on the ends induces a spontaneous unfolding order 
for the individual structural elements, which is a function of their 
relative stabilities and the topology of the structure. 
In contrast, the nanopore prescribes a linear order along the sequence, 
and unfolds an RNA molecule much as enzymes such as the ribosome do 
in cells. This difference suggests that the two approaches can also 
yield different types of information about the molecule under study. 
As demonstrated by Onoa {\it et al.} \cite{Onoa03}, clever use of the 
pulling on the ends approach can reveal detailed information on the 
(un)folding pathway of an RNA molecule with known structure. 
However, when the structure of an RNA molecule is unknown, pulling on 
the ends can provide, by itself, little information beyond a count of the 
number of structural elements that unfold separately \cite{Onoa03,Gerland03}. 
In the following we therefore focus on the question of how much 
structural information may in principle be obtained with the nanopore 
approach. 

Let us suppose that we were able to observe the trajectories $m(t)$ of the 
molecules during the translocation process. 
We could then assign a position within the sequence to each arrest 
during translocation. 
Since an arrest is caused by a kinetic barrier, i.e. a stem trapped at 
the entrance to the pore, we could thereby identify the positions of 
the stems in the structure. 
Such information can indeed be sufficient to reconstruct almost the 
entire basepairing pattern of a molecule, as we demonstrate explicitly 
using an example below. 
If the translocation dynamics is in the strongly driven limit (B) where the 
structure on the {\it cis} side is essentially frozen, then the reconstructed 
structure would correspond to the initial structure of the molecule before 
translocation. 
We concentrate on this limit in the following, including a discussion of its 
attainability. 
However, it may be noteworthy that in the slow translocation limit (A) one 
would also obtain useful structural information, namely on the average 
structure of the molecule (with respect to the thermodynamic ensemble of all 
structures \cite{McCaskill90}). 
As long as the molecule is `well-designed' this average structure will be 
dominated by the ground-state, i.e. the minimum binding free energy 
structure\footnote{The worst case for the purpose of structure determination 
corresponds to the regime where the typical timescale for the translocation 
of say a single hairpin is comparable to the timescale for structural 
rearrangements involving the formation of new stems: in this case, the 
structure on the {\it cis} side may relax after a stem is unzipped, so 
that one would oberve only the signatures of the relaxed structure rather 
than the original structure. 
This regime should be avoided by a proper choice of the driving force 
and the friction coefficient of the pore (R.~Bundschuh and U.~Gerland, 
to be published).}. 

How could one possibly observe the trajectories $m(t)$ during translocation? 
For the purpose of structure determination, we will need $m(t)$ with a 
spatial resolution below the typical length of a stem in an RNA 
structure (5--10 basepairs). 
This may be achievable through a refinement of the current nanopore 
technology, such that careful analysis of the ionic current allows a 
count (or even sequencing) of the bases that have passed the 
pore \cite{Akeson99,Meller00}. 
With artificial solid-state pores \cite{Li01,Storm03} it is also 
conceivable to use a tunneling current through leads within the 
membrane as a probe to count (or sequence) the bases as they pass 
through the pore. 
Here, we explore yet another option, namely pulling the molecule 
mechanically through the pore, with a device that can record 
force-extension curves, e.g. an atomic force microscope or optical 
tweezers. The explicit discussion of this case with an exemplary 
RNA sequence serves us to gauge the more general capability 
of nanopores as single-molecule tools for the study of biopolymer 
folding. 

\noindent {\bf Quantitative aspects.} 
Mechanical unfolding of a biopolymer yields characteristic sawtooth-shaped 
signatures in the force-extension curve (FEC) indicating the opening of 
structural elements within the molecule, see e.g. \cite{Rief97,Onoa03}. 
{From} the relative positions of these sawteeth one can determine length 
changes within the molecule with an extremely high resolution of about 1~nm. 
In the usual setup where the molecule is unfolded by pulling on its ends, 
such length changes can only be used to infer the `stored length' of a 
structural element, but not its precise position along the backbone of the 
molecule, cf.~Fig.~\ref{comparison}. 
In contrast, for mechanical pulling through a pore, the relative positions of 
the resulting sawteeth will correspond directly to the relative positions 
of the structural elements in the sequence\footnote{The absolute position 
can be inferred by adding a known structural element, 
e.g. a strong C-G hairpin, to one end of the RNA, which can then function as 
a reference point.}. 
One conceivable way to prepare the initial condition where an RNA molecule 
is almost entirely on the {\it cis} side, with one end threaded through the 
pore and attached to a pulling device on the {\it trans} side, 
is to start with an attached molecule on the 
{\it trans} side and to apply a voltage pulse across the pore that 
suffices to drive the molecule as far as possible to the {\it cis} side. 

To apply our general model to the particular case of mechanical pulling 
in the strongly driven (fast pulling) limit, we need to specify the form of 
the three terms in the free energy landscape (\ref{general_FE}). 
The second term, i.e. the binding free energy on the {\it trans} side, may 
be set to zero, 
\begin{equation}
  \cF_{trans}(m)=0 \;,
\end{equation}
since the reformation of structure after 
translocation is suppressed at high tensions in the RNA single 
strand\footnote{For instance, Liphardt {\it et al.} \cite{Liphardt01} 
observed refolding rates for a single hairpin around 1~s$^{-1}$ at the 
unfolding force $f_{1/2}\approx 14$~pN. At a pulling speed of say 
$1\,\mu$m/s, the translocation of an RNA molecule with a thousand bases 
would therefore be terminated before refolding of a structural element 
on the {\it trans} side occurs.}. 
The third term, $\cF_{\rm ext}(m)$, describes the effect of the mechanical 
stress on the RNA, which stretches the single-stranded {\it trans} part of 
the molecule. 
The elastic response of this single-strand may be modeled by a freely 
jointed chain (FJC) polymer model. 
Assuming for simplicity a constant pulling speed $v$, the third term then 
takes the form 
\begin{equation} 
  \label{F_ext}
  \cF_{\rm ext}(m)=\cF_{{\rm FJC+spring}}(v\cdot t;N-m)\;. 
\end{equation}
Here, the function $\cF_{{\rm FJC+spring}}(R_t;n)$ denotes the combined 
free energy of a single-stranded RNA of $n$ bases in series with a linear 
spring, stretched to a total extension $R_t=v\cdot t$ \cite{Gerland03}. 
(The linear spring takes into account the stiffness of the force-measuring 
device, see the Appendix for details.) 
By assumption, the first term, $\cF_{cis}(m)$, represents the binding free 
energy of the remaining part of the initial structure on the {\it cis} side. 
$\cF_{cis}(m)$ can be calculated for any initial structure, based on the 
free energy rules for RNA secondary structure \cite{Walter94} with a 
natural extension for pseudoknotted structures, see the Appendix. 
Our assumption of a frozen structure on the {\it cis} side is most likely 
an oversimplification for realistic pulling speeds, since small 
fluctuations in the secondary structure are known to occur already on 
timescales on the order of tens of microseconds \cite{Bonnet98}. 
However, since the pore pulling approach is sensitive only to stem 
positions, we expect that it is unaffected by small fluctuations and 
sensitive only to major rearrangements which significantly change the 
secondary structure. 
Such rearrangements are typically slow, sometimes even on the 
timescale of hours \cite{Pan98,Zhuang00}. 

\begin{figure}[tb]
  \includegraphics[width=8cm]{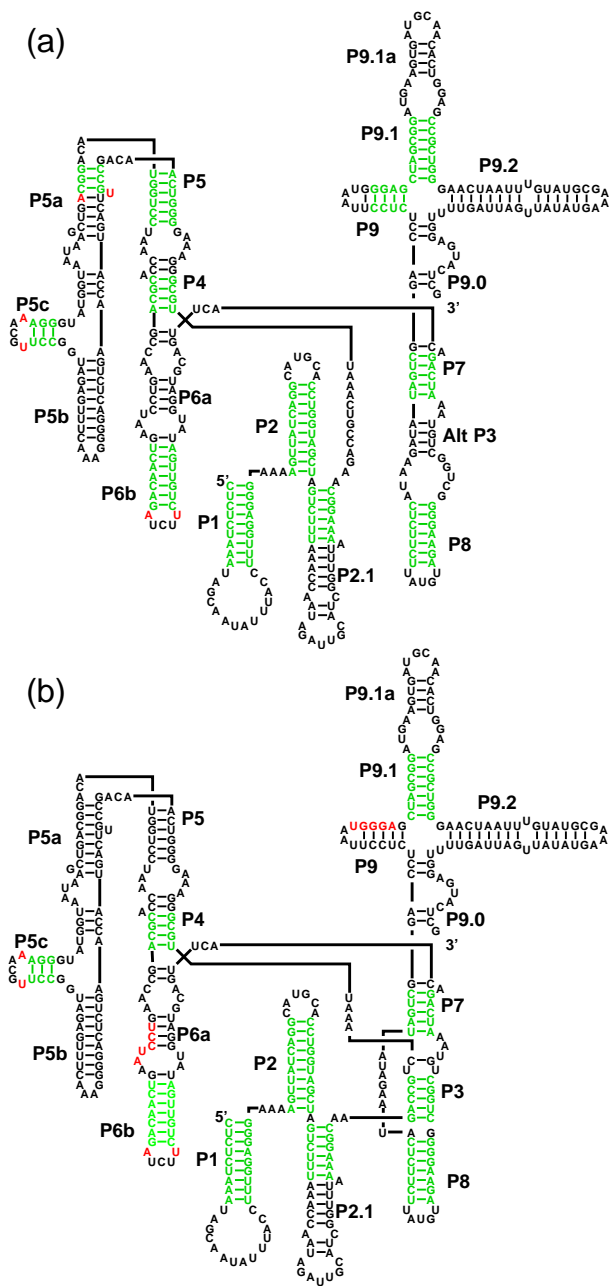}
  \caption{Secondary structure of the {\it Tetrahymena thermophila} Group I 
  intron: (a) Long-lived folding intermediate \cite{Pan98}. 
  (b) Native state with pseudoknot. 
  The basepairs shown in green are correctly reconstructed from the 
  force-extension curves, see Fig.~\ref{fig_quickpull}, using the 
  procedure described in the main text, while the bases shown in red are 
  involved in incorrect basepair predictions (the procedure yields no 
  prediction for the bases shown in black); see also 
  Fig.~\ref{FIGreconstruction}.} 
  \label{figstruct} 
\end{figure}

\vfill\noindent {\bf Reconstruction of secondary structures.} 
To illustrate the problem and the method, 
we use an exemplary RNA, the well-studied self-splicing intron of 
{\it Tetrahymena thermophila} \cite{Cech93} with a sequence of 419 
bases (Genbank \# V01416). 
In its correctly folded active state, the basepairing pattern of this 
ribozyme contains a pseudoknot (see Fig.~\ref{figstruct}a), while its 
best characterized long-lived folding intermediate \cite{Pan98,Zhuang02} 
has a known alternative structure without pseudoknot 
(see Fig.~\ref{figstruct}b) \cite{Pan98}. 
We will investigate whether one can in principle use the 
pulling-through-a-pore approach not only to discriminate between these two 
different conformations in individual molecules, but also to reconstruct 
both structures from the FEC's. 

\begin{figure}[t]
  \includegraphics[width=8.5cm]{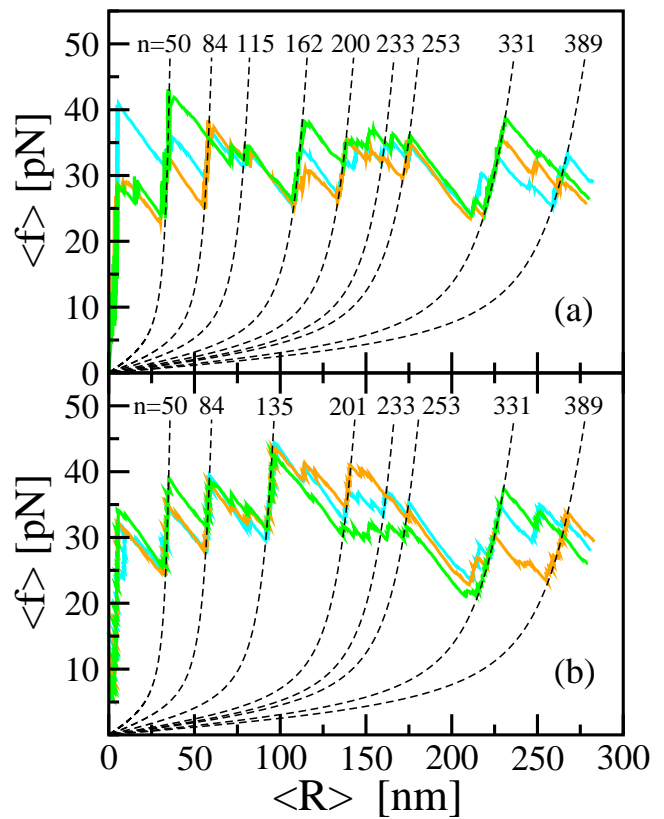}
  \caption{Force-extension traces (solid lines) as obtained with our 
  stochastic model for mechanical pulling through a nanopore. 
  (a) and (b) each show three different runs with the same initial conditions 
  (and pulling speed of $v=0.1$~nm/time step) for the structures in 
  Fig.~\ref{figstruct}(a) and (b), respectively. The force $\langle f\rangle$ 
  and extension $\langle R\rangle$ are calculated using Eq.~(\ref{force}) 
  in the Appendix. 
  The dashed lines are freely jointed chain FEC's whose lengths 
  are fitted to some of the positions that correspond to translocation 
  arrests.}
  \label{fig_quickpull} 
\end{figure}

To obtain FEC's for these structures, we performed Monte-Carlo simulations 
of the stochastic process defined by 
Eqs.~(\ref{hopping}--\ref{hopping_rate}), and used Eq.~(\ref{force}) from 
the Appendix to calculate the force and extension time traces. 
We performed all calculations at the same pulling speed 
($v=0.1$~nm/time step, which roughly corresponds to 10~$\mu$m/s given 
typical values for $k_0$, see above), 
and the same stiffness of the force-measuring device ($\lambda=0.5$~pN/nm). 
Fig.~\ref{fig_quickpull} displays three such FEC's (solid lines) for the 
non-pseudoknotted structure of Fig.~\ref{figstruct}b, corresponding to 
unzipping from the 3' end. 
These FEC's show the sawtooth-like behavior which is characteristic for the 
sequential opening of structural elements (a very similar behavior was 
observed in the experiments of Onoa {\it et al.} \cite{Onoa03} where the 
molecule was rapidly unzipped by pulling on its ends). 
The rising parts of the sawteeth correspond to stretching of single strand 
on the {\it trans} side as a stacked region is ``trapped'' in 
front of the pore on the {\it cis} side. 
When a stacked region opens, some single strand is freed 
to pass the pore, which leads to relaxation of the tension and causes the 
downstrokes in the FEC's. 
Note that the FEC's do not share all of their sawteeth, which reflects 
the importance of thermal fluctuations for this type of single molecule 
experiments (this property is manifest also in the experiment of Onoa 
{\it et al.} \cite{Onoa03}). 

The most relevant information contained in the FEC's are the positions 
of the translocation arrests, during which the required force for the 
opening of basepairs is built up. 
To extract these positions, we use FEC's of freely jointed chains with 
different lengths: 
The dashed lines in Fig.~\ref{fig_quickpull} show some examples of 
such FEC's where the chain length $n$ coincides with the length of the 
RNA single strand on the {\it trans} side during such an arrest. 
With an automated procedure described in the Appendix we obtain all of these 
positions (above a threshold for the duration of an arrest). 

\begin{figure}[t]
  \includegraphics[width=8.6cm]{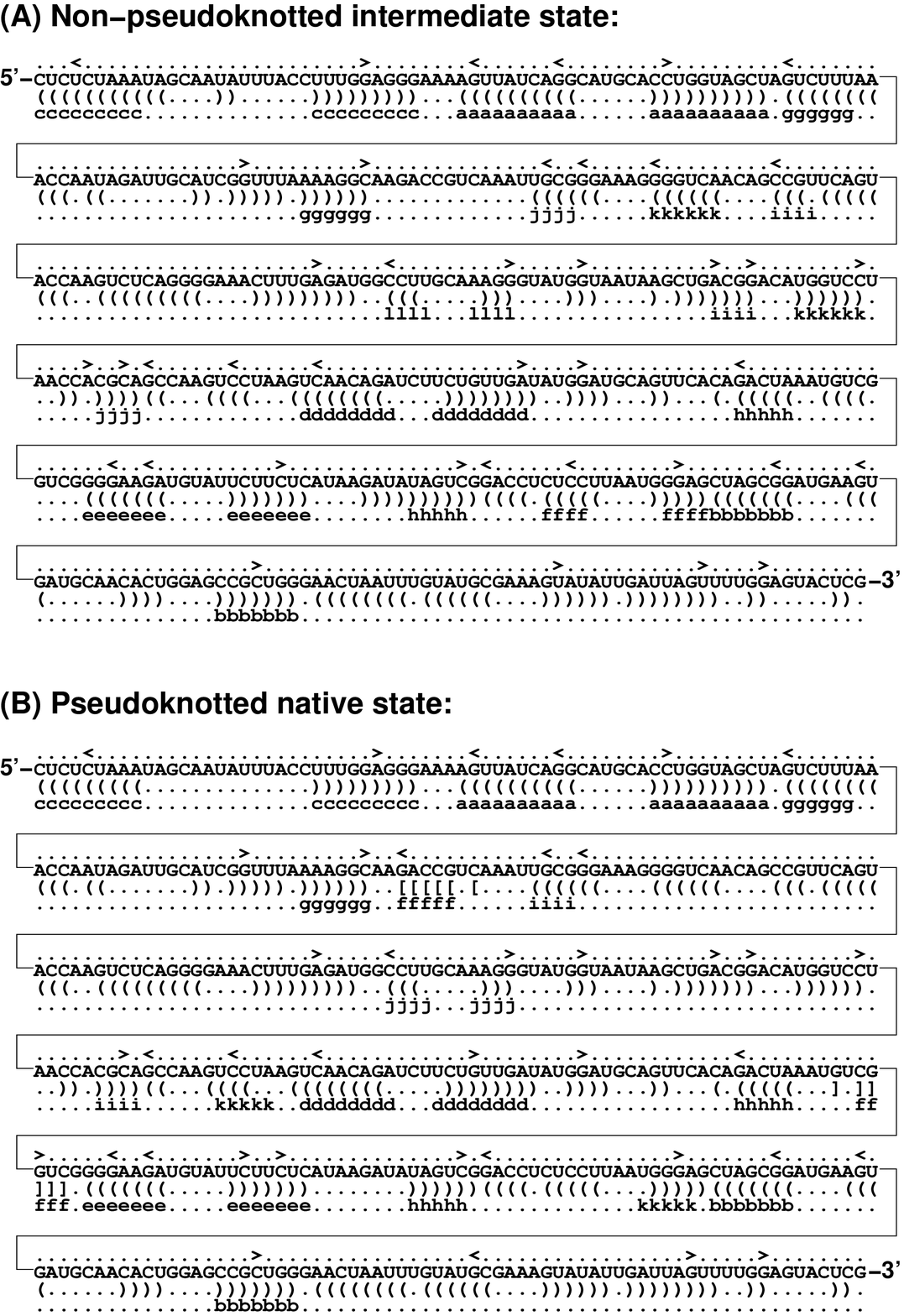}
  \caption{Reconstruction of the basepairing pattern from the FEC's. 
  First row: parentheses extracted from the FEC's, which indicate the position 
  of basepaired regions. Second row: RNA sequence. 
  Third row: parentheses indicating the basepairs in the full structures 
  shown in Fig.~\ref{figstruct}. 
  Fourth row: stems predicted from the parentheses in the first row by sequence 
  alignment. See main text for details.}
  \label{FIGreconstruction} 
\end{figure}

Since the bases around the position of an arrest are very likely 
basepaired with another segment of the sequence further to the 5' end, we 
represent this information by a closing angular bracket, `$\rangle$', above 
that position in the RNA sequence (written from 5' to 3'), see 
Fig.~\ref{FIGreconstruction}. 
Of course, the molecule can also be pulled through the pore in the other 
direction, i.e. from the 5' end. 
This yields information on the positions of segments that have downstream 
binding partners. 
The same procedure then leads to the opening brackets, `$\langle$', also 
shown in Fig.~\ref{FIGreconstruction}. 

Bracket representations are a widely used short hand notation for RNA 
secondary structures. For the structures in Fig.~\ref{figstruct}, such a
representation is shown in the third row of Fig.~\ref{FIGreconstruction}. 
Note that two types of brackets have to be used for the pseudoknotted native 
structure, in order to make the association between opening and closing 
brackets unambiguous. 
We observe that the angular brackets extracted from the FEC's can be viewed as 
an incomplete bracket representation of the RNA secondary structure. 
Can we complete it using only the given sequence of the RNA molecule? 

This task is a sequence alignment problem, which consists of matching each 
opening (closing) bracket with an associated downstream (upstream) binding 
sequence. Several circumstances conspire to make this, somewhat surprisingly, 
a nontrivial problem: 
(i) stems, i.e. contiguous basepaired regions, are usually short, typically 
5--10 basepairs, 
(ii) structural elements often lead to a different number of angular brackets 
in the two pulling directions, i.e. not every opening bracket has a 
corresponding closing bracket and {\it vice versa}, and 
(iii) sequence segments containing several U's have many possible binding 
partners, since U's can pair with A's and G's. 

To overcome this problem, we developed a probabilistic sequence alignment 
algorithm (see Appendix), which identifies the most likely set of stems that 
is consistent with all angular brackets and where all paired sequence segments 
contain at least one angular bracket on each side. 
The output of this algorithm is shown in the fourth rows of 
Fig.~\ref{FIGreconstruction}, where lower case letters indicate paired sequence 
segments and the alphabetic order represents the confidence level (confidence 
is largest for `a'). 
In Fig.~\ref{figstruct} the bases involved in this reconstructed set of stems 
are colored, with green (red) indicating (in)correct basepairing. 
We observe that the two different basepairing patterns (for the same sequence) 
are clearly distinguished and the large scale secondary structure is captured 
in both cases. 
In particular, the pseudoknot in the native structure is correctly 
identified. The only incorrectly predicted stem is the least significant one 
(`k') in the pseudoknotted structure. 

While these results seem satisfactory as a proof of principle, we stress that 
our reconstruction algorithm can certainly be improved upon, e.g. by 
allowing for mismatches in longer stems, which should help to fill in many of 
the missed basepairs. Also, one could make use of the known basepairing 
energies in the reconstruction.

\section*{Discussion and Outlook} 
Our theoretical study has led us to a simple coarse-grained model, 
Eqs.~(\ref{general_FE}-\ref{hopping_rate}), for the translocation of 
structured polynucleotides, which is applicable in the two opposite limits 
of very slow and very rapid translocation. 
This model is a useful starting point for a more detailed description that 
remains valid in the entire parameter regime. 
Here, we have applied the model to demonstrate that the physics of the 
translocation process can in principle be exploited to use nanopores 
for secondary structure determination (including pseudoknots) on the 
single-molecule level. 
Indeed, the nanopore technique would be a useful addition to the 
existing repertoire of structure determination methods: 
RNA secondary structure can be predicted computationally to some extent 
\cite{Zuker81,McCaskill90,Vienna94} based on experimentally determined 
free energy rules \cite{Walter94}, however this approach 
is unreliable for RNA molecules exceeding $\sim 100$ bases and cannot 
take pseudoknots properly into account. 
Including pseudoknots, which are often crucial to the function of RNA 
enzymes \cite{Pan98,Wadkins99}, is not only computationally expensive 
\cite{Rivas99,Isambert00}, but is also limited by a lack of experimental 
information on the corresponding binding free energies. 
Experimentally, X-ray crystallography \cite{Cate96} or NMR 
\cite{Colmenarejo99} provide detailed structures, but these techniques are 
cumbersome and limited to small molecules or isolated domains of larger RNAs. 
Structural information for larger RNAs can currently only be obtained 
from comparative sequence analysis \cite{Gutell92}, which requires large sets 
of homologous RNA sequences, or from indirect biochemical 
methods \cite{Pan98}. 

Throughout this paper, we have focused on basepairing only, which is 
permissible under ionic conditions that disfavor tertiary interactions, 
e.g. low sodium and no magnesium. 
However, once the translocation of a molecule is well characterized under 
these conditions, it becomes interesting to switch to the native ionic 
conditions and examine the effects of tertiary interactions. 
Generally, one can expect more cooperativity in the presence of tertiary 
interactions, i.e. larger domains will open in a single step, as observed 
by Onoa {\it et al.} \cite{Onoa03}. 
This suggests a hierarchical approach to structure determination with 
nanopores: first unzip under low ionic conditions to obtain the secondary 
structure, and then repeat in the presence of magnesium to identify how the 
secondary structure elements are grouped into larger tertiary structure 
domains (such as the P4-P6 domain in the {\it Tetrahymena} ribozyme). 
It is worthwhile to stress the advantage of RNA as a model system to 
separately study the effect of secondary and tertiary structure. 
In contrast, the secondary structure of proteins is not stable in the absence 
of tertiary structure, and hence one may expect that single-domain proteins 
will unfold and translocate across a pore in a single step. 

Nanopores could in principle also be used to probe the kinetics of 
large-scale secondary structure rearrangements in single-molecules. 
For instance, it would be useful to attach larger objects to both ends of 
a molecule that is already threaded through the pore, allowing the same 
molecule to be driven forth and back through the pore, over and over again. 
By varying the time interval between successive reversals of the driving 
force, one could then probe structural relaxation over a broad range of 
time scales. 
More generally, nanopores may emerge as a new tool to probe intra- and 
inter-molecular interactions in single biomolecules. 
For instance, one could probe the biophysics of combined binding and 
folding in the context of RNA-protein interactions. 


\noindent {\bf Acknowledgments.} 
We are grateful to D.~Branton, S. Ling, J.~Liphardt, and D.~Lubensky 
for stimulating discussions. 
This research was supported by the National Science Foundation through 
grant No. 0211308, 0216576, and 0225630.

\section*{Appendix} 
\noindent {\bf Calculation of free energy landscape.} 
Given a secondary structure of the molecule, we obtain $\cF_{cis}(m)$ by 
eliminating all basepairs involving the terminal $N\!-\!m$ bases, and 
calculating the binding free energy of the remaining structure according to 
the free energy rules for RNA secondary structure \cite{Walter94}. 
[We take the free energy parameters as supplied with the Vienna RNA package 
(version 1.3.1) at room temperature $T=25^oC$. The salt concentrations at 
which these parameters were measured are $[\rm{Na}^+]=1M$ and 
$[\rm{Mg}^{++}]=0M$.] 
For pseudoknotted structures, the free energy rules currently include no 
prescription, however the following extension appears reasonable: we 
first eliminate basepairs in stems that give rise to the pseudoknot(s) and 
calculate the free energy of the remaining structure according to the standard 
rules. We then add the free energies of the eliminated stems separately, 
including the free energy for the loops created by these stems, again according 
to the standard free energy rules (however, the bases in these loops that are 
involved in other stems are removed before calculating the loop free energy). 

The {\it trans} part of the molecule is tethered at both ends, by the pore 
and the pulling device, respectively. 
The pulling device can be described by a linear spring, while the 
configurational entropy of the RNA single strand can be modeled by a 
freely jointed chain (FJC) with extensible segments. 
[For the few bases that are inside the pore, we neglect the effect of the 
confinement on the entropy.] 
We denote by $R_t$ the total extension of the {\it trans} part in series 
with the linear spring. 
The free energy (\ref{F_ext}) can then be expressed in terms of the total 
end-to-end distance distribution $W_{\rm FJC+spring}$,
\begin{displaymath}
  \cF_{{\rm FJC+spring}}(R_t;n)=-k_BT \log W_{\rm FJC+spring}(R_t;n) \;, 
\end{displaymath}
which can in turn be written as the convolution of the individual end-to-end 
distance distributions of the FJC and the spring \cite{Gerland03}, 
\begin{displaymath}
  W_{\rm FJC+spring}(R_t;n)=\!\int\limits_0^{\infty}\!\!{\rm d}R\;
  W_{\rm FJC}(R;n)\,W_{\rm spring}(R_t\!-\!R)\:. 
\end{displaymath}
Here, 
$W_{\rm spring}(R_s)=\exp(-\beta\lambda{R_s}^2/2)/\sqrt{2\pi/\beta\lambda}$, 
where $\lambda$ denotes the inherent stiffness of the pulling device. 
We calculate the end-to-end distance distribution of the freely jointed chain, 
$W_{\rm FJC}(R;n)$, as described previously \cite{Gerland01}. 
The polymer parameters we use were obtained from a fit \cite{Mezard01} to 
FEC's of single-stranded DNA \cite{Maier00} (base-to-base length $0.7\,$nm, 
Kuhn length $1.9\,$nm, and stretch modulus $815\,$pN), since we are unaware of 
corresponding data for the chemically very similar RNA.

\vfill\noindent {\bf Calculation of FEC's.} 
We obtain several trajectories $m(t)$ with a Monte Carlo simulation of 
Eqs.~(\ref{general_FE}--\ref{hopping_rate}) with $m(0)=N$, $R_t(0)=0$ as 
initial condition and incrementing $R_t$ at the constant rate $v$. 
The simulation is stopped when all bases have translocated ($m=0$). 
{From} the time trace $m(t)$, we calculate the force-extension curve $f(R)$ 
using 
\begin{equation}
\label{force}
  \langle f\rangle = \frac{\partial}{\partial R_t} 
  \cF_{\rm FJC+spring}(R_t\!=\!vt;\,N\!-\!m(t)) 
\end{equation}
and $\langle R\rangle = vt - \langle f\rangle/\lambda$. 
Here, $\langle f\rangle$ and $\langle R\rangle$ are both thermal averages 
over the polymer and spring 
degrees of freedom at fixed total extension $R_t$ and fixed basepairing 
pattern.

\vfill\noindent {\bf Extraction of parentheses positions from FEC.} 
For every point on a FEC, we determine the length $n$ of the freely jointed 
chain whose FEC passes closest to the point (using the polymer parameters for 
single-stranded RNA as given above). 
We take a histogram of the resulting lengths $n$ over three independent FEC's 
for each structure. 
In this histogram, the lengths $n$ that correspond to start positions of 
stably basepaired regions appear as peaks, since the length of single-stranded 
RNA on the {\it trans} side remains approximately constant while the force 
required to unzip the basepairs builds up. 
[A similar procedure was applied in Ref.~\cite{Koch02} to identify the 
positions of proteins bound to double-stranded DNA as it is being unzipped.] 
We keep all $n$-values where the histogram exceeds a threshold of $30$ counts 
(a count is made every Monte Carlo time-step).
Since thermal noise makes the molecule fluctuate back and forth by a 
few bases while the force is building up for the next stem to open, we pick 
out of each contiguous stretch in the remaining $n$-values only the largest. 
Finally, we increment the extracted $n$-values by one and mark the 
corresponding position in the sequence with a parenthesis.

\vfill\noindent {\bf Reconstruction of basepairing pattern.} 
The FEC's do not reveal which opening and closing parentheses are paired with 
each other. However, given the sequence of the RNA, we can match the 
parentheses by sequence complementarity. 
[To keep the number of false basepair predictions to a minimum, we consider 
only stems where we have at least one parenthesis at each end.] 
Here, we summarize the essential steps in our sequence alignment algorithm, 
while a detailed presentation and characterization will be given elsewhere 
(R.~Bundschuh and U.~Gerland, to be published): 
First, we find all possible gapless local alignments between a subsequence 
containing a parenthesis and subsequences to the open side of the parenthesis, 
using the scoring scheme 2 for GC, 1 for AU, and 0 for GU. 
We keep only those alignments with a score larger than 5 and where the matching 
sequence segment also contains a matching parenthesis. 
We consider the remaining alignments as possible stems in the secondary 
structure. To pick the most likely set of mutually consistent stems, we 
assign an alignment E-value to each stem \cite{Karlin90}. 
We then iteratively include the most likely stem into the structure prediction, 
and remove all other stems it excludes due to overlapping basepairs from the 
list of allowed stems. 

%





%
\end{document}